\documentclass{PoS}

\title{Automatic O$(a)$ improvement for twisted-mass QCD}

\ShortTitle{Automatic O$(a)$ improvement for twisted-mass QCD}

\author{\speaker{Sinya Aoki}\\
        Graduate School of Pure and Applied Sciences, University of
        Tsukuba, Tsukuba, Ibaraki 305-8571, Japan, 
	and \\ 
	Riken BNL Research Center, Brookhaven National Laboratory, Upton,
	NY 11973, USA \\
        E-mail: \email{saoki@het.ph.tsukuba.ac.jp}}
\author{Oliver B\"ar\\
        Institute of Physics, Humboldt University Berlin, Newtonstrasse
	15, 12489 Berlin, Germany \\
        E-mail: \email{obaer@physik.hu-berlin.de}}

\abstract{We present a condition for automatic O$(a)$ improvement in
twisted mass lattice QCD, using symmetries of the Symanzik effective
theory. If the continuum part of the Symanzik effective theory is invariant
under a particular transformation, named $T_1$ in this report,
scaling violations of all quantities 
invariant under $T_1$ transformation are even in the lattice spacing $a$. 
On the other hand, quantities non-invariant under $T_1$ vanish in the
continuum limit with 
odd powers in $a$.
We prove this statement even for the massive case without using
the equation of motion.
We also consider a few different criteria for the $T_1$ invariant condition
in lattice theories and discuss ambiguities of the lattice condition for
O$(a)$ improvement.
}
\FullConference{XXIVth International Symposium on Lattice Field Theory\\
                July 23-28, 2006\\
                Tucson, Arizona, USA}

\begin{document}

\section{Introduction}
It becomes more and more apparent that twisted mass Lattice QCD
(tmLQCD) \cite{Frezzotti:2001ea,Frezzotti:2000nk} is a promising
formulation to approach the chiral limit of QCD, despite the fact
that the flavor symmetry is explicitly broken.
A twisted mass protects the Wilson-Dirac operator against small
eigenvalues and therefore solves the problem
of exceptional configurations \cite{Aoki:1989,Aoki:1990}, 
thus making numerical simulations with small quark
masses feasible \cite{Aoki:1992}. 
This numerical advantage of tmLQCD is supplemented by the property of
automatic O($a$) improvement
\cite{Frezzotti:2003ni}. 
For a recent 
review of these and some more results in twisted mass LQCD see Ref.\ 
\cite{Shindler:2005vj}.

The so-called ``maximal twist'' condition, required for the proof of 
automatic O($a$) improvement \cite{Frezzotti:2003ni}, however,
causes some confusions.
For example, it has been suggested \cite{Frezzotti:2003ni} that
maximal twist can be achieved by setting the bare untwisted mass
to the critical quark mass of the Wilson fermion where the pion mass
vanishes (we call this  choice ``the pion mass definition'' in the following).
However, it has been pointed out \cite{Aoki:2004ta} that
this choice does not lead to automatic O$(a)$ improvement.
Indeed,  terms linear in $a$ and with fractional powers of $a$
are predicted by Wilson Chiral Perturbation Theory (WChPT) for very
small twisted quark masses. On the other hand, automatic O($a$)
improvement is expected to hold if the critical mass is defined
through the partially conserved axial vector Ward identity quark mass
(PCAC mass definition).

In this report we present an explicit condition based on symmetries of
the Symanzik theory, 
required for 
automatic O$(a)$ improvement, and prove that scaling violations
of all quantities which have non-zero values in the continuum limit 
are even in $a$ (O$(a)$ improvement). The detailed proof of this
statement has already been published in Ref.\ \cite{Aoki:2006}. 
Therefore, in this report we try to avoid unnecessary rigorousness
in our proof and stress the
mechanism which leads to O$(a)$ improvement in twisted mass QCD.

\section{Condition and proof for automatic O$(a)$ improvement}
We first give our statement, which will be proven in this report.
Twisted mass QCD with a certain condition leads to automatic O$(a)$
improvement, which means that operators as well as the action are
automatically $O(a)$ improved without any
improvement coefficients. 
This holds true even in the massive case and without the explicit use of the equations of motion.
More explicitly, all
scaling violations of non-zero physical quantities are even in $a$, while
quantities which vanish in the continuum limit have only odd powers
in $a$.

\subsection{Main idea of the proof}
The twisted mass lattice QCD action for the 2-flavor theory is given by
$
S_{\rm tm} = S_G + S_F,
$
where $S_G$ is the gauge action and
\begin{eqnarray}
S_F &=& \sum_{x,\mu} \bar \psi_L (x) \frac{1}{2}\left[ \gamma_\mu
(\nabla_\mu^+ + \nabla_\mu^-)\psi_L
-ar \nabla_\mu^+ \nabla_\mu^-\psi_L\right](x) +
\sum_x\bar\psi_L (x) M_0 e^{i\theta_{0} \gamma_5\tau^3} \psi_L (x)
\label{eq:lattice_action}
\end{eqnarray}
is the 2-flavor Wilson fermion action with a twisted  mass term,
where $M_{0}$ and $\theta_{0}$ denote the bare mass and bare
twist angle. It is also
customary to write
$
M_0 e^{i\theta_{0} \gamma_5\tau^3} = m_0 + i \mu_{0}\gamma_5\tau^3,
$
using the bare untwisted mass $m_{0}$ and the bare twisted mass $\mu_{0}$.

This action is invariant under the following global transformations:
(1) U(1)$\otimes$U(1) vector symmetry,
$
\psi_L \rightarrow e^{i (\alpha_0 + \alpha_3\tau^3)}\psi_L$,\,
$\bar\psi_L \rightarrow \bar\psi_L e^{-i (\alpha_0 + \alpha_3\tau^3)}$ .
This transformation is part of the U(2) flavor symmetry of the
untwisted theory.
(2) Extended parity symmetry
$
P_F^{1,2}:\  \psi_L (x) \rightarrow  \tau^{1,2} \gamma_4\psi_L(Px)$,
$
\bar\psi_L(x) \rightarrow  \bar\psi_L(Px)\gamma_4 \tau^{1,2}
$
where $P$ is the parity transformation.
Alternatively, one can also augment $P$ with a sign change of the
twisted mass term $\mu_{0}$,
$
\tilde{P}  = P \times [\mu_{0} \rightarrow -\mu_{0}]$,
which is also a symmetry of the action.
(3) Standard charge conjugation symmetry, as in the untwisted theory.

The lattice theory can be described by an effective continuum theory
(the Symanzik theory),
whose effective action is restricted by
locality and  the symmetries of the underlying lattice theory. Taking
into account the symmetries listed above one finds
$
S_{\rm eff} = S_0 + a S_1 + a^{2} S_{2} + \cdots$,
where the first two terms are given as
\begin{eqnarray}
S_{0} &=& S_{0,\rm gauge} + \int d^4x\, \bar\psi(x) \left[ \gamma_
{\mu}D_{\mu} +
M_R e^{i\theta\gamma_5\tau^3}\right]\psi(x),\label{eq:symanzik_S0}\\
S_{1} & = & C_1
\int d^4x\, \bar\psi(x) \sigma_{\mu\nu} F_{\mu\nu}(x) \psi (x).
\label{eq:symanzik_S1}
\end{eqnarray}
$S_{0,\rm gauge}$ denotes the standard continuum gauge field action
in terms of the gauge field tensor $F_{\mu\nu}$. The second term in
$S_{0}$ is the continuum twisted mass fermion action. 
It is worth mentioning that there
is no ``twisted'' Pauli term $\bar\psi \gamma_5\tau^3 \sigma_{\mu\nu}
F_{\mu\nu}
\psi$ present in $S_{1}$, since such a term violates the symmetry 
$\tilde P$.

In addition to the effective action we have to specify the direction
of the chiral condensate, since chiral symmetry is spontaneously
broken. From the fact that the direction of the chiral condensate is
completely controlled by the direction of the symmetry breaking
external field
(i.e. the quark mass) in the continuum theory, we can take
\begin{eqnarray}
\langle \bar\psi_\alpha^i\psi_\beta^j\rangle_{S_0}
&=& \frac{v(M_R)}{8}\left[ e^{-i\theta\gamma_5\tau^3}\right
]_{\beta\alpha}^{ji},
\label{eq:vev0}
\end{eqnarray}
where $\lim_{M_R\rightarrow 0} \lim_{V \rightarrow\infty} v(M_R)\not=
0$.
Here the vacuum expectation value (VEV) is defined with respect to
the continuum action $S_0$.
To say it differently, the VEV (\ref{eq:vev0}) defines the twist
angle $\theta$ in the Symanzik theory.

We now want to argue
that the choice $\theta = \pi/2$ (or $-\pi/2$) corresponds to
``maximal twist''. In terms of the mass parameters this is equivalent
to $M_{R} = \mu_{R}$ and $m_{R}=0$. In this case the action and the
VEVs become
\begin{eqnarray}
S_0 &=&S_{0, \rm gauge}+ \int d^4x\ \bar\psi (x)\left[
\gamma_{\mu} D_{\mu}+ i M_R  \gamma_5\tau^3 \right] \psi (x), 
\end{eqnarray}
$\langle \bar\psi \psi\rangle_{S_0} = 0$,
$
\langle \bar\psi i\gamma_5\tau^3 \psi\rangle_{S_0} =  v(M_R)$ .
It is easy to check that $S_0$,
the continuum part of the effective action
is invariant under
$\psi \rightarrow e^{i w\gamma_5\tau^{1,2}}\psi$, 
$\bar\psi \rightarrow \bar\psi e^{i w\gamma_5\tau^{1,2}}$,
and therefore also under the $Z_2$ subgroup $T_1$ of this continuous
transformation,
defined by
$T_1 \psi = i\gamma_5 \tau^1 \psi$, 
$T_1\bar\psi =  \bar\psi i\gamma_5 \tau^1$.
Since $T_1^2 =1$ in the space of fermion number conserving operators,
which contain equal numbers of $\psi$ and $\bar\psi$,
the eigenvalues of $T_1$ are $1$ ($T_{1}$-even) or $-1$ ($T_{1}$-odd).
The crucial observation is that the VEVs 
$\langle \bar\psi \psi\rangle$ and
$\langle \bar\psi i\gamma_5\tau^3 \psi\rangle$
are also invariant
under this transformation. The $T_1$ symmetry 
is {\em not}  spontaneously broken, hence it
is an exact symmetry of the continuum theory.
The O$(a)$ term
$a S_1 = a C_1\int d^4 x\
\bar\psi(x) \sigma_{\mu\nu} F_{\mu\nu}(x) \psi (x)$,
on the other hand, is odd under $T_1$.
Therefore, non-vanishing physical observables, which
must be even under $T_1$, can not have an O$(a)$ contribution, since
the O$(a)$ term is odd under $T_1$ and therefore must vanish
identically.
This is automatic O$(a)$ improvement at ``maximal twist''.
Note that non-invariant, i.e.\ $T_1$-odd quantities, which vanish in
the continuum limit, can have O$(a)$ contributions.

The above argument gives just the main idea of our proof for
automatic O$(a)$ improvement, and we will give a detailed proof in
the next subsection. However,
one of the most important points of our analysis is that the
condition for automatic O$(a)$ improvement is 
the invariance of theory under $T_1$ transformation, or more generally
its continuous version, which
corresponds to a part of the exact vector symmetry in
continuum QCD at ``maximal twist''.
One might even say that the $T_1$ invariance is more fundamental for automatic
$O(a)$ improvement than the notion of ``maximal twist'', which is one of the
consequences of $T_1$ invariance.

\subsection{General proof}
Let us consider an arbitrary  multi-local lattice operator 
$ O_{\rm lat}^{tp,d}( \{x\} )$,
where $\{x\} $ represents $x_1,x_2, \cdots, x_n$,  $d$ is the canonical
dimension
of the operator,  $t=0,1$ and $p=0,1$ denote the transformation
properties under $T_1$ and $P$:
\begin{eqnarray}
T_1 :\  { O}_{\rm lat}^{tp,d} (\{x\}) &\rightarrow& (-1)^t{ O}
_{\rm lat}^{tp,d} (\{x\}) , \quad
P : \ { O}_{\rm lat}^{tp,d} (\{\vec{x}, t\}) \rightarrow (-1)^p
{ O}_{\rm lat}^{tp,d} (\{ -\vec{x},t \}) .
\end{eqnarray}
Here we do not include the dimension coming from powers of the quark
mass  in the canonical dimension $d$ of the operators. 

The lattice operator ${O}_{\rm lat}^{tp,d}$ corresponds to a sum of
continuum operators ${O}^{t_n p_n,n}$  ($n$: non-negative
integer) in the Symanzik theory as
\begin{eqnarray}
{O}_{\rm lat}^{tp,d} 
&=&\sum_{n=d}^\infty a^{n-d} \sum_{t_n,p_n} c_{t_n p_n,n}^
{tp,d} {O}^{t_n p_n,n}  ,
\label{eq:OP_expand}
\end{eqnarray}
where $n$ is the canonical dimension of the continuum operator 
${O}^{t_np_n,n}$
which consists of  $\bar\psi$, $\psi$, $A_\mu$ and $D_\mu$ only
without any mass parameters, and
\begin{eqnarray}
T_1 : \  {O}^{t_np_n,n} (\{x\})&\rightarrow & (-1)^{t_n}
{O}^{t_np_n,n}(\{x\}),\quad
P : \  { O}^{t_np_n,n}(\{ \vec{x},t \}) \rightarrow  (-1)^
{p_n} {O}^{t_np_n,n} (\{ -\vec{x},t \}),
\end{eqnarray}
with $t_n, p_n =0,1$. 
To have a total dimension $d$ in the expansion in 
Eq.\ (\ref{eq:OP_expand}),
the coefficients $c_{t_np_n,n}^{tp,d}$ must be dimensionless.
Here, to make an argument simpler, we consider the lattice operator, whose
power divergences can be subtracted without spoiling
our proof\cite{Aoki:2006}.

The following selection rules among these operators
are crucial for our proof of automatic O$(a)$ improvement:
\begin{eqnarray}
t  + p+d &=& t_n  + p_n +n\quad {\rm mod} (2) , \quad
p+ \#\mu_0 = p_n + (\#\mu_0)_n \quad {\rm mod} (2),
\label{eq:rule}
\end{eqnarray}
where  $\#\mu_0$ and $(\#\mu_0)_n$ denote the numbers of $\mu_0$'s in $
{O}_{\rm lat}^{tp,d}$ and
$c_{t_np_n,n}^{tp,d}$, respectively.
The second equality can be easily proven by the invariance of the
lattice action
(\ref{eq:lattice_action}) under the $\tilde P =  P \times [\mu_0
\rightarrow -\mu_0 ]$ transformation.
To prove the first equality, we introduce
the following transformation:
\begin{eqnarray}
{\cal D}_d^1 : \left\{
\begin{array}{ccc}
U_\mu(x) & \rightarrow & U^\dagger_\mu(-x-a\mu) \\
\Bigl(A_\mu (x) &\rightarrow & - A_\mu(-x) \Bigr)\\
\psi(x) &\rightarrow &\left( e^{i\pi\tau_1}\right)^{3/2}\psi(-x) \\
\bar\psi(x) &\rightarrow & \bar\psi(-x) \left(e^{i\pi\tau_1}\right)^
{3/2} \\
\end{array}
\right. ,
\label{eq:Dd1}
\end{eqnarray}
which is a modified version of the transformation ${\cal D}_d$
introduced in Ref.~\cite{Frezzotti:2003ni}.
Since it is easy to show that
the lattice action (\ref{eq:lattice_action}) is invariant under $T_1
\times {\cal D}_d^1$, in addition to the invariance under $P_F^1$,
the lattice action is invariant under $T_1 \times {\cal D}_d^1\times
P_F^1$.
On the other hand,
we can easily see that
${\cal D}_d^1\times P_F^1$ counts the canonical dimension times the parity
of the operator as
\begin{eqnarray}
{\cal D}_d^1 \times P_F^1 : \ {O}_{\rm lat}^{tp,d}( \{\vec{x},t\})
&\rightarrow & (-1)^{d+p}
    { O}_{\rm lat}^{tp,d}( \{\vec{x},-t\}), \\
{\cal D}_d^1 \times P_F^1 : \ { O}^{t_np_n,n}( \{\vec{x},t\}) &
\rightarrow & (-1)^{n+ p_n}
{ O}^{t_np_n,n}( \{\vec{x},-t\}) .
\end{eqnarray}
Therefore, the invariance of the action under $T_1\times {\cal D}_d^1
\times P_F^1$ implies the first selection rule.

Let us show how these selection rules are used to determine the
structure of operators in the Symanzik theory.
We first consider the Symanzik expansion of the lattice action $S_{\rm
tm}$:
\begin{eqnarray}
S_{\rm tm} &=& \sum_{n=0}^\infty \left[ a^{2n}(S^{00,2n} + \mu_0 a\,
				  S^{11,2n})
+a^{2n-1}(\mu_0 a\, S^{01,2n-1}+ S^{10,2n-1})
\right] \\
&=& S_0^{0} + m S_{-1}^1 + \sum_{n=1}^\infty
\left[ a^{2n} S_{2n}^0+a^{2n-1}S_{2n-1}^1
\right],
\label{eq:symanzik_action}
\end{eqnarray}
where $S^{t_n p_n,n}$ denotes the action in the Symanzik theory whose
canonical dimension and transformation properties under $T_1$ and $P$
are $(d_n, t_n, n)$. Here we pull out the $\mu a$ factor from terms
which have odd powers in $\mu_0 a$. Therefore remaining factors always have
even powers in $\mu_0 a$.
To derive the first equality we use the selection
rules such that
$0 = n+t_n +p_n \,{\rm mod} (2)$ 
$0 = p_n + (\# \mu_0)_n \, {\rm mod} (2)$
since the lattice action satisfies $d+t+p=0$ and $p+\# \mu_0=0$. 
In the
second equality we define
\begin{eqnarray}
S_{2n}^0 &=& S^{00,2n} + \mu_0 S^{01,2n-1}, \quad
m S_{-1}^1 =S^{10,-1}/a,\
S_{2n-1}^1 = S^{10,2n-1} + \mu_0 S^{11,2n-2}(n\ge 1), 
\end{eqnarray} 
and the superscript 0 or 1 denotes the transformation property under 
$T_1$ as evident from the above definition.
Similarly we have
\begin{eqnarray}
{ O}_{{\rm lat},d}^0 &=& {O}_d^0 +\sum_{n=1}^\infty 
\left[ a^{2n}{ O}_{d+2n}^0 + 
+a^{2n-1}{ O}_{d+2n-1}^1 \right] ,\\
{O}_{{\rm lat},d}^1 &=& O_d^1 +\sum_{n=1}^\infty 
\left[ a^{2n}{ O}_{d+2n}^{1} + 
+a^{2n-1}{ O}_{d+2n-1}^0 \right] ,
\end{eqnarray}
for the multi-local operator with the canonical dimension $d$,
where again the superscript 0 or 1 denotes the transformation property
under $T_1$.

Now we can specify the condition for automatic $O(a)$ improvement:
It is stated that the continuum part of the action 
(\ref{eq:symanzik_action}) is invariant under $T_1$.
This condition leads to $m=0$, so that the continuum part of the action
is given solely by $S_0^0$. We will consider the scaling behaviour of
the vacuum expectation value of an arbitrary multi-local operator,
$\langle {O}_{{\rm lat},d}^t  (\{x\} ) \rangle$.
For this purpose we define
\begin{equation}
e^{S_{\rm tm}} = e^{S_0^0}\exp \left\{\sum_{n=1}^\infty
\left[
a^{2n} S_{2n}^0 + a^{2n-1} S_{2n-1}^1 \right] \right\}
\equiv e^{S_0^0} \sum_{n=0}^\infty a^n S^{(n)} ,
\label{DefExpSEff}
\end{equation}
where we define
$a^n S^{(n)}$ to be the sum of the $a^n$ terms in
eq.\ (\ref{DefExpSEff}). 
For example, the first few terms are given as
$S^{(0)} = 1$, 
$S^{(1)} = S_1^1$ , 
and $S^{(2)} = S_2^0 + {(S_1^1)^2}/{2!}$ .
Under the $T_1$ transformation, they behave as
$T_1 :\  S^{(n)}\rightarrow  (-1)^n S^{(n)}$ .
By expanding both action and operator, we have
\begin{eqnarray}
\langle O_{\rm lat}^t (\{x\} ) \rangle
&=& \sum_{n=0}^\infty a^n \langle O_n^{t_n}(\{x\} )
 \rangle_{S_{\rm tm}}
=\sum_{n=l=0}^\infty a^{n+l} \langle O_n^{t_n}(\{x\} )
 S^{(l)}\rangle_{S_0^0}
\end{eqnarray}
where $t_n=n+t$ mod (2). In the second line the $T_1$ invariance tells us that
$\langle O_n^{t_n} (\{x\} ) S^{(l)}\rangle_{S_0^0}=0$ unless 
$t_n+l = t+ n +l =0$ mod (2). Therefore we have
\begin{eqnarray}
\langle O_{\rm lat}^t (\{x\} ) \rangle
&=& \sum_{s=0}^\infty a^{2s+t}\sum_{n=0}^{2s+t} \langle O_{n}^{t_n}
(\{x\} ) S^{(2s+t-n)}\rangle_{S_0^0} ,
\end{eqnarray}
from which we derive
\begin{eqnarray}
\langle O_{\rm lat}^0 (\{x\} ) \rangle
&=& \langle O_0^0 (\{x\} ) \rangle_{S_0^0}+O(a^2) + O(a^4)+\cdots
\\
\langle O_{\rm lat}^1 (\{x\} ) \rangle
&=& O(a) + O(a^3)+O(a^5)+\cdots .
\end{eqnarray}
This proves our statement that the scaling violation of all
$T_1$ invariant operators, which have non-zero VEV in the continuum limit,
are even in $a$, while that of $T_1$ non-invariant operators,
whose VEV vanish in the continuum limit, are odd in $a$.
This is true for non-zero $\mu_0$ and does not require the use of the equation
of motion.

\subsection{Condition for O$(a)$ improvement in the lattice theory}
In the Symanzik theory, the condition for O$(a)$ improvement is uniquely 
defined by the condition that an arbitrary $T_1$ non-invariant operator
${ O}^{t=1\, p, d}$ has a vanishing expectation value. Provided this
condition is fulfilled, the expectation values of all
$T_1$ non-invariant operators vanish. Hence the particular choice for
${ O}^{1p, d}$ is irrelevant, and in that sense the condition 
is unique.
In the lattice theory, however,  the condition
defined by
$\langle  {O}_{\rm lat}^{1p,d} \rangle = 0$
depends on the choice of the operator ${ O}_{\rm lat}^{1p,d}$, and is 
therefore not unique. In terms of the Symanzik theory, for this condition
to be satisfied,
an equation
\[
a F_0 (a^2, ma, \mu_0) + m F_1(a^2, ma,\mu_0) = 0,
\]
where $F_{0,1}$ are some functions of $a^2$, $ma$ and $\mu_0$,
must be fulfilled by tuning the untwisted mass $m$. Since this equation
is invariant under $(m,a)\rightarrow (-m,-a)$, the solution has the form that
$m = a f(a^2,\mu_0)$ under the assumption that the solution to the
equation is unique. If one takes a different lattice operator to
define the $T_1$ invariant condition,
the solution is given by $m^\prime = a
f^\prime(a^2,\mu)$. Therefore the difference between two definitions is
O$(a)$: $m-m^\prime = a(f-f^\prime)$.
Note that a solution $m$ in  general depends on $\mu_0$, 
inherited from the $\mu_0$ dependence of $F_{0,1}$.

Let us consider some examples for the condition in the lattice theory.
A simple one is given by
$\langle \left(\bar\psi\psi\right)_{\rm lat}\rangle = 0$ .
Unfortunately,  this definition is not very useful in practice, since
the subtraction of power divergences necessary for $\langle\bar\psi
\psi\rangle$
prevents a reliable determination of this VEV in the lattice theory.
Instead one may take ${O}_{\rm lat}(x,y) = A_\mu^a (x) P^a (y)$ or
${O}_{\rm lat}(x,y) = \partial_\mu A_\mu^a (x) P^a (y)$ ($a=1,2$),
as was done in  Refs.\
\cite{Bietenholz:2004wv,Abdel-Rehim:2005gz,Jansen:2005gf}:
\begin{eqnarray}
\label{LattCond2}
\langle A_\mu^a (x) P^a (y)\rangle &=& 0 \quad \mbox{ or } \quad
\langle \partial_{\mu}  A_\mu^a(x) P^a (y) \rangle = 0 ,
\label{eq:parity_con}
\end{eqnarray}
where $A_\mu^a$ and $P^a$ denote the axial vector current and pseudo
scalar density, respectively.
Yet another choice is \cite{Sharpe:2004ny}
\[
\langle A_\mu^3 (x) P^3 (y)\rangle = 0.
\]
Depending on the choice for the axial vector current, either the
local or the conserved one,
the conditions (\ref{LattCond2}) lead to a different definition for
maximal twist. However, the difference will be again of O$(a)$.

We close this section with a final comment.
Any condition for O$(a)$ improvement in the lattice theory determines
a value for the 
bare untwisted mass 
$m_{0}$ as a function of the bare twisted
mass $\mu_0$. It has been suggested to tune the untwisted mass to its
critical value $m_{0}=m_{\rm cr}$ where the pion mass vanishes in the
untwisted theory. However, this condition is not related to  $T_1$
invariance. For example,
contributions from excited states violate eq.\ (\ref{eq:parity_con})
even at $m_\pi =0$.
Consequently, the pion mass definition does not correspond to
automatic O$(a)$ improvement according to the  $T_{1}$ invariance 
condition.

\section{Conclusion}
In this paper we gave a comprehensive proof for automatic O$(a)$
improvement in twisted mass lattice QCD.
The most important observation is that
a precise definition for O$(a)$ improvement is described by the
symmetry in the continuum theory. If the continuum part of the Symanzik
theory is invariant under $T_1$ transformation, scaling violations for
all quantities are shown to be even powers in $a$,
as long as they are invariant under the $T_1$ transformation.
Non-invariant quantities, on the other hand, vanish as odd powers in
$a$.

\vskip 0.5cm

This work is supported in part by the Grants-in-Aid for
Scientific Research from the Ministry of Education,
Culture, Sports, Science and Technology.
(Nos. 13135204, 15204015, 15540251, 16028201),
O. B.\ is supported in part by the University of Tsukuba Research
Project.

\end{document}